# Post-crisis Strategies:
# Antifragility Principles as Catalysts for Urban Evolution Towards Sustainability

*Joseph Uguet[a], Nicola Tollin[b] & Jordi Morató[a]*

[a]*UNESCO Chair on Sustainability, Universitat Politècnica de Catalunya-BarcelonaTech, Spain*

[b]*Southern Denmark University, Denmark ,*

## Abstract

Urban crises reveal the true essence of cities: their ability to either withstand disorder or collapse under its pressure. This article explores how antifragility principles can transforms urban disruption into levers for reinforcement and innovation. While resilience seeks to restore a lost balance, antifragility goes further: it pushes cities to improve through shocks. Across a critical analysis of post-crisis strategies and the identification of fifteen fundamental theoretical principles, this work proposes a new framework, structuring a proactive and evolutionary approach to urban development. Medellín, Singapore and Fukushima already illustrate this dynamic, showing that adversity can catalyse profound transformations. By integrating institutional flexibility, strategic diversity and self-organization, antifragility poses itself as an alternative to the limits of resilience. Can this model really redefine the way cities adapt to crises? This article paves the way for a decisive reflection to rethink urban planning in an uncertain world.

## Introduction

Far from temporary disruptions, today's urban crises mark structural inflection points, prompting cities to reconsider both their responses and their development pathways. These disasters with a dizzying socio-economic cost (COVID-19, Fukushima, Haiti, Katrina, Valencia, etc.) brutally expose the flaws of standard models that manage at best to restore their initial functionality as quickly as possible (Holling, 1973). But in a context where extreme events are increasingly shaping the new frame of reference (Taleb 2007), can we be satisfied with maintaining the status quo?

Faced with these limitations, a bolder aspiration is emerging: the "Antifragility" concept (Taleb, 2012). Where resilience tends to return to an original state after a shock (Holling, 1973), this paradigm invites us to thrive through disruption. The adage "what does not kill me makes me stronger" (Nietzsche, 1888), precisely embodies the dynamic of learning and reinforcement through adversity proposed by this concept. Applied to the urban fabric, it offers the opportunity to transform challenges into levers of innovation and evolution.

Some cities have demonstrated that remarkable reversals are possible by turning their crises into powerful drivers of transformation. Medellín, once plagued by violence, has become a model of social urbanism and inclusion; Singapore, faced with water scarcity, now excels in water management; Fukushima, despite the nuclear disaster of 2011, has established itself as a world leader in nuclear decontamination and deconstruction. These trajectories are reminiscent of the evolutionary dynamics described by Lamarck (1809), where environmental adversity pushes adaptation. These territories have thus been able to convert their hardships into levers of transformation, equipping themselves with smart infrastructure, strengthening citizen engagement and adopting an adaptive and sustainable vision of development (Jacobs, 1961).

This article examines this conceptual transition from resistance to (proactive) transformation. The critical analysis of standard crisis management strategies will thus highlight their limits in the face of contemporary challenges. It will then identify the fundamental principles that shape an antifragile urban fabric and finally propose a renewed conceptual framework, adapted to post-crisis territories and allowing to recode the sustainable city in an uncertain world.

**Review of Existing Urban Crises Adaptation Strategies**

*Role of urban crises as pivots of transformation: Fragility, Robustness and Resilience*

Urban crises reveal the true essence of cities: their ability to overcome disruptions or to collapse under the pressure it brings. These extreme events mark breaking points, where the city oscillates between fragility and renewal. They test the limits of existing systems and expose their flaws while paving the way for potential transformations (Davoudi, 2012; UN-Habitat, 2021).

Urban crisis response strategies reflect the ability to anticipate, absorb and transform disruption into opportunity or fail and collapse (UN-Habitat, 2021). These tests expose the intrinsic vulnerabilities of the systems in place by challenging their ability to overcome them and adapt (Davoudi, 2012).

Three fundamental states stand out in this dynamic: **fragility**, characterizing an inability to absorb shocks (De Bruijn, 2019; Platje, 2015); **robustness**, ensuring structural stability in the face of disturbances (Taleb 2012); and **resilience**, embodying the ability to absorb stress and bounce back from extreme events (Holling, 1973; Platje, 2015).

**Fragility and systemic traps, a multidimensional state**: Urban fragility stems from the complex interplay of economic, social, environmental and institutional factors, which exacerbate vulnerabilities to crises. It manifests itself in markers such as poverty, unemployment, limited access to essential services, and pollution, compounded by uncontrolled urbanization and weak governance, severely undermine cities' ability to recover (Platje, 2015). These dynamics illustrate a multidimensional reality of fragility, described by Muggah and Savage (2012) and the OECD (2015), who identify five interrelated dimensions: violence, justice, institutions, economic foundations, and resilience.

The case of Damascus is a sad example of this trajectory, where the destruction caused by the Syrian conflict since 2011, combined with excessive centralization and a lack of resources, shows how fragility can become a chronic state.

Multiple and recurrent crises reinforce these vulnerabilities by locking urban systems into "self-reinforcing traps", where each disruption amplifies existing structural failures (Munoz et al., 2022). These negative cycles, fuelled by obsolete infrastructures and inadequate management, freeze urban systems in trajectories of stagnation or even collapse (AFD, 2016).

The cases of Port-au-Prince and Beirut illustrate this mechanism. In 2010, the earthquake in Port-au-Prince exposed the inability of institutions to respond effectively, prolonging a lasting socio-economic paralysis. The Beirut port explosion in 2020 crystallized decades of mismanagement, crippling the city's vital functions. These examples show that without appropriate strategies, crises transform fragilities into permanent states, limiting any prospect of reconstruction or evolution.

**Robustness, a static resistance:** Robust cities rely on strong infrastructure and preventative measures to maintain stability in the face of disruptions. This defensive approach aims to limit the immediate impacts of crises without initiating profound structural transformations. It is necessary in contexts where the recurrence of crises requires an effective short-term response by favouring operational continuity, often to the detriment of the adaptability necessary to deal with complex or unforeseen crises (Taleb, 2012; OECD, 2015; AFD, 2016).

Mumbai and Havana offer telling examples of robustness in action. In Mumbai, frequent flooding has led to massive investments in drainage systems and preventive policies, reducing material losses. Similarly, Havana, thanks to its early warning systems and community mobilization, manages to maintain relative stability in the face of recurrent hurricanes. These experiments show that robustness is an effective first line of defence against frequent shocks.

However, this model quickly shows its limits. By maintaining a status quo, it inhibits the learning and adaptability needed in the face of systemic or large-scale disruptions (Muggah & Savage, 2012; Munoz et al., 2022). Far from strengthening urban resilience, this rigidity institutionalizes a form of latent vulnerability, exposing cities to more severe disruptions when crises exceed the expected response capacities.

Robustness is thus distinguished by immediate stability, but insufficient when challenges become evolving (de Bruijn, 2019). While it provides a necessary foundation, it must be complemented by strategies that integrate adaptability and transformation, the only paths to sustainable responses to contemporary upheavals.

**Resilience, limited capacity for adaptation and transformation:** The adaptability of cities in the face of crises depends on their ability to withstand disruptions while maintaining their essential functions. Inspired by the work of Holling (1973), this dynamic is based on a structural plasticity that allows urban systems to reorganize themselves after a shock. Anchored in the agenda of the *Sendai Framework* (UN, 2015), it promotes risk anticipation and consolidation of critical infrastructure to mitigate the impact of disasters.

Some cities have integrated these principles into development strategies that transcend risk management. In Rotterdam, the design of the "Water Squares" illustrates this change. These multifunctional spaces, both temporary reservoirs and living spaces, transform the hydrological constraint into a lever for sustainable development (Ilgen et al, 2019). This "Living with Water" approach inspired similar interventions in New Orleans after Hurricane Katrina (Fields et al., 2016). Tokyo takes this integration even further by combining earthquake-resistant infrastructure, inclusive governance and nature-based solutions (Tokyo Resilient Project, TMG, 2023). Far from a defensive posture, this strategy combines anticipation, modularity and innovation to anchor the city in a logic of proactive adaptation.

While urban resilience has established itself as a dominant frame of reference in contemporary urban policies, its application reveals deep contradictions and structural impasses. Behind an apparent conceptual neutrality, it struggles to respond to the power imbalances that shape urban vulnerability. Too often, it is reduced to a technical pragmatism, obsessed with the restoration of the existing rather than with a structural transformation of the inequalities that make cities vulnerable in the first place (Davoudi, 2012).

This approach, although effective in limited contexts, is based on a linear anticipation of risks, which is ill-suited to the dynamics of uncertainty and systemic crises (Blečić & Cecchini, 2020). It produces standardized solutions, disconnected from local realities, and accentuates the fragmentation of urban responses (Davoudi, 2012; Blečić & Cecchini, 2020). Rather than acknowledging the interconnectedness of crises—ecological, economic, and social—it encloses each disruption within siloed sectoral frameworks, unable to capture the deep interactions that shape urban trajectories (Munoz et al., 2022).

Resilience strategies also suffer from a top-down approach that marginalizes the dynamics of self-organization and endogenous adaptation (dos Passos et al., 2018). This lack of inclusiveness limits the scope of initiatives and widens the gap between institutional design and local ownership. Rather than fostering innovation, technocratic governance often preserves outdated systems and slows down the strategic recombination needed for sustainable transformation (Roggema, 2021).

However, an urban model that does not progress under the effect of crises is condemned to suffer their recurrence. Taleb (2012) highlights this inertia and highlights the need for a paradigm reversal: rather than clinging to a logic of conservation and stabilization, cities must learn to take advantage of disorder to amplify their capacity for adaptation. Disruptions should not be seen as hazards to be neutralized, but as catalysts for transformation, likely to reveal new urban opportunities and recompose governance structures.

Resilience often becomes an approach that focuses only on city services (water, energy, mobility, etc.) while overlooking residents' capacities and the social and cultural dimensions, among others. This service-centric view reflects a technocratic bias that obscures community agency, social capital, and cultural practices that shape recovery and adaptation, and it can exacerbate inequalities. A people-centered, justice-oriented resilience framework should systematically integrate residents' capabilities, social infrastructure, and cultural dimensions−using participatory diagnostics, metrics for cohesion, trust, and inclusion, and investments that strengthen local institutions and livelihoods.

Finally, the dominant resilience strategies continue to be marked by a Western bias, based on central planning and heavy infrastructure. These approaches obscure vernacular forms of adaptation and neglect local dynamics of self-organization, particularly in cities in the Global

South (Wahba, 2021). This exclusion weakens the scope of the policies put in place and hinders the emergence of responses that are truly adapted to vulnerable contexts.

Behind the apparent consensus on the need to strengthen the capacity of cities to overcome crises, the criticisms converge: conceptual imprecision (Davoudi, 2012), inadequacy to systemic crises (Munoz et al., 2022), lack of truly inclusive approaches (dos Passos et al., 2018), institutional inertia (Roggema, 2021), rejection of transformation as an adaptive lever (Taleb, 2012) and marginalization of local strategies (Wahba, 2021).

Absorbing crises is no longer enough, they must be converted into engines of transformation. The city of the twenty-first century must stop preserving its fragility under the guise of adaptation and switch to logics of learning, self-organization and adaptive innovation, it must tend towards antifragility.

**Antifragility: From Theoretical Foundations to its Urban Expression**

*Foundations of antifragility*

Antifragility has its roots in centuries-old reflections on the ability of systems to evolve in the face of crises. While helping to establish biology as a discipline, Lamarck (1809) described the gradual adaptation of organisms to environmental constraints, a process that reflects reinforcement through adversity. Then Grassé's (1959) research on the concept of stigmergy demonstrated how local interactions can produce optimized global structures, suggesting self-organization pathways applicable to complex ecosystems. They will inspire technologies such as blockchains and can naturally be transposed to the urban fabric. It was Holling (1973) who extended this perspective by showing that certain systems, subjected to shocks, reorganize themselves to better cope with uncertainty.

More than two centuries after Lamarck, Taleb (2012) formalized these intuitions into a conceptual triad: (i) fragility—systems that break under volatility (in the face of chaos); (ii) robustness—systems that withstand shocks without changing; and (iii) antifragility—systems that improve through exposure to disturbances.

The foundations of antifragility lie in key principles such as diversity, redundancy and optionality. These ideas, although explored in various fields — from finance to biology — converge towards a universal capacity of systems to transform disorder into opportunity.

Since Taleb's foundational work, the paradigm has been expanded by dozens of academic contributions across diverse fields. This study focuses on a curated panel of 33 authors (Appendix 1). Figure 1 provides a non-exhaustive overview of their principal contributions to the paradigm. For instance, Johnson (2013) developed a methodological framework to assess antifragility in complex systems. Dos Passos et al. (2018) and de Bruijn et al. (2019) highlighted the roles of self-learning, adaptability, and feedback dynamics as drivers of systemic evolution.

Meerow et al. (2016) emphasized functional diversity and adaptive flexibility within urban planning. In turn, Shafique (2016) placed optionality and functional fluidity at the core of antifragile urbanism, proposing spatial configurations capable of shifting use according to changing conditions.

For more than a decade, these interdisciplinary perspectives have shown how antifragility, initially theoretical, has gradually been transformed into a strategic tool applicable to various contexts and in particular to that of the urban fabric.

To better understand cities' possible trajectories in the face of crisis, Figure 2 illustrates distinct dynamics arising from revealed vulnerabilities and chosen responses, incorporating the notion of antifragility. Four theoretical urban trajectories—**fragile, robust, resilient**, and **antifragile**—are compared under two successive shocks, revealing increasing divergence among their dynamics and highlighting differences in their capacity to react, recover, and evolve.

Before the impact, all cities were on the same level, reflecting an apparent lack of differentiation between their preparations, and thus hiding their latent weaknesses. With each shock, these vulnerabilities are exposed, placing the city in a critical zone where three main trajectories emerge. Some cities enter a downward spiral of collapse, unable to mobilize the necessary resources or initiate a rebound. They sink into chaos, marked by a worsening of structural failures. Others achieve only a limited rebound: despite conventional recovery mechanisms, they manage to return to a level no higher than their original state. Finally, a transformative rebound dynamic, characteristic of antifragile systems, enables certain cities to go beyond mere recovery by reinventing their priorities and integrating structural innovations. This process propels them to a higher level of performance after each impact.

Phases of revelation—when fragilities are exposed—and incubation—when initial responses are shaped—play a pivotal role in setting trajectories. Preparedness and rapid post-shock action determine whether a city declines, stabilizes, or transforms.

These differences become particularly pronounced after several disruptions, where only antifragile cities demonstrate a true ability to thrive in instability, widening a significant gap with other models. This Figure 2 highlights the importance of strategic choices and innovation dynamics in building a sustainable urban future.

To illustrate this, in 2005 Hurricane Katrina dramatically exposed New Orleans' structural weaknesses, including the failure of levees and their disproportionate impact on marginalized neighbourhoods. This event catalysed debates on environmental justice and equitable management of infrastructure, highlighting the importance of implementing new strategies such as "Living with Water" (Fields et al., 2016).

The COVID-19 pandemic has redefined urban dynamics globally. By challenging standard models, it has encouraged the emergence of adaptive solutions such as reversible public spaces, local health infrastructure and soft mobility (UN-Habitat, 2021).

These reflections highlight the potential of extreme events to trigger systemic transformations and reconfigure urban priorities. They create tipping points where urban models can be rethought to better adapt to future uncertainties. The fundamental question remains how to harness these critical moments to redefine urban trajectories and build more scalable and sustainable systems.

*Fundamental theoretical principles and their dynamics in urban antifragility*

Amid the complex interactions that shape cities, antifragility has emerged over the past decade as a conceptual compass for rethinking post-crisis dynamics. It proposes a transformative reading framework where disturbances become catalysts for urban evolution. This section presents the process of extracting and synthesizing the key theoretical principles that emerge from fundamental research, which has greatly enriched the academic literature over the past decade.

**Extraction methodology**: In order to identify the foundations of urban antifragility, a heuristic 3 steps process was adopted. Building on an analytical classification of the key references (cf. Figure 1), the literature review enabled the extraction of **core concepts**, which were structured into five recurring **thematic patterns**. From these, three sub-groups of interrelated logics were identified, whose convergence laid the foundation for the formulation of the fifteen **fundamental theoretical principles** underpinning the paradigm of urban antifragility.

**Concepts → Patterns → Theoretical Principles**

*Figure 1. Heatmap – Intensity of Principal Authors' Contributions (Panel 33)*

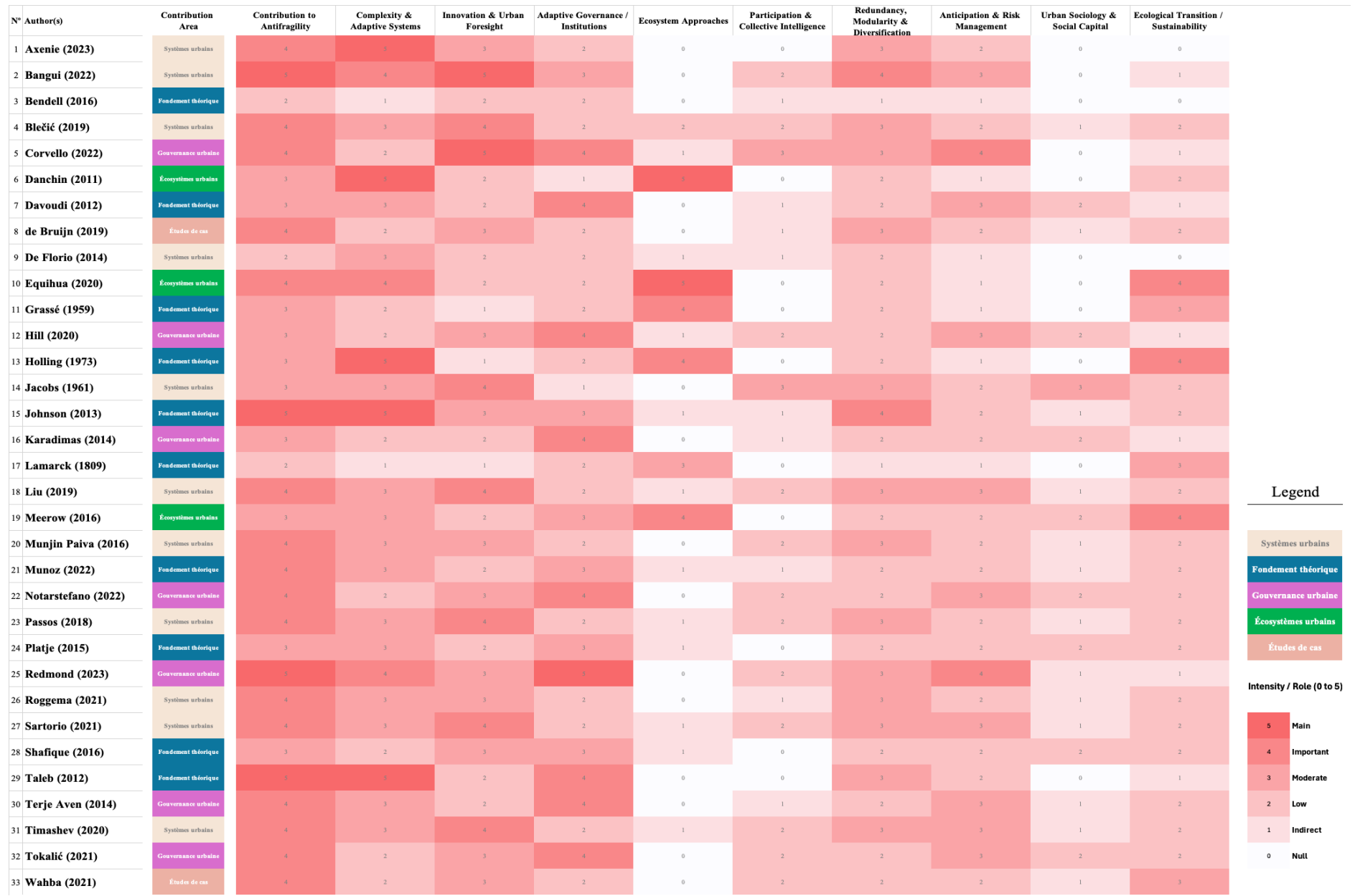

*Author: J.Uguet*

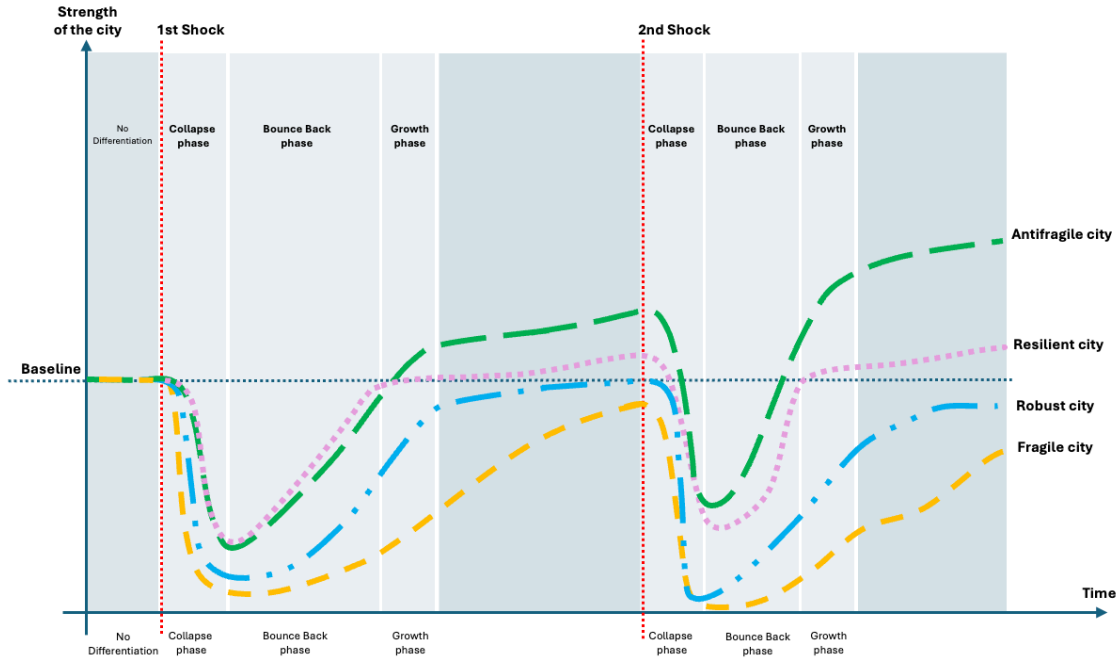

*Figure 2. Urban Trajectories in the Face of Crises: From Fragility to Antifragility. Author: J.Uguet*

The results of this in-depth analysis are based on the contributions of fifteen major authors, whose work, rooted in the study of the urban fabric, structures this theoretical corpus. The panel was chosen for research that bridges conceptual frameworks to empirical analyses of urban systems and governance. Together with case studies, these investigations show how cities turn crises into opportunities for transformation. Table 1 provides the authors and an overview of their contributions.

Each reference and their contributions have been positioned according to five main axes, with the associated key concepts (cf. Appendix 1). This segmentation has made it possible to bring out thematic convergences and to gradually structure the reference links.

This initial classification enables a **rapid identification of key conceptual clusters, which converge around five recurring thematic patterns**: Adaptation and Self-Organization, Proactive Innovation, Diversity and Strategic Redundancy, Governance and Resource Mobilization, and Proactive Prevention and Sustainable Resilience. These motifs—frequently cited across the antifragility literature—have provided an evolutionary foundation, further refined through the analysis of conceptual nexuses in the selected authors' work. Their distribution is illustrated in Figure 3.

*Table 1. Key References' panel and main contributions - Author: J.Uguet*

| Author(s), Date | Contribution Area | Key Concepts (Keywords) |
|---|---|---|
| **Bangui et al., 2022** | Urban systems | Blockchain, AI, responsiveness, adaptability |
| **Blečić et al., 2020** | Urban systems | Anti-fragile planning, flexibility, modularity |
| **De Bruijn, 2019** | Case Studies | Redundancy, experimentation, dynamical systems |
| **Equihua et al., 2020** | Urban ecosystems | Biodiversity, information theory, adaptation |
| **Johnson, 2013** | Theoretical basis | Analysis, measurement, antifragility |
| **Munjin Paiva, 2016** | Urban systems | Architectural design, flexibility, adaptability |
| **Notarstefano, 2022** | Urban governance | Empowerment, participatory governance, active sustainability |
| **Passos et al., 2018** | Urban systems | ICT, self-organization, self-learning |
| **Platje, 2015** | Theoretical basis | Institutional fragility, vulnerability indicators |
| **Redmond et al., 2023** | Urban governance | AntifragiCity, predictive tools, rapid mobilization |
| **Roggema, 2021** | Urban systems | Spatial redundancy, flexibility, climate resilience |
| **Sartorio et al., 2021** | Urban systems | Urban morphology, modularity, adaptive transmission |
| **Taleb, 2012** | Theoretical basis | Fragile-robust-antifragile triad, via negativa, optionality |
| **Timashev, 2020** | Urban systems | Supra-resilience, critical infrastructure, interconnection |
| **Wahba, 2021** | Case Studies | Inclusive reconstruction, post-crisis, infrastructure |

*Figure 3. Matrix of distribution of key dimensions of antifragility by key authors- Author: J.Uguet*

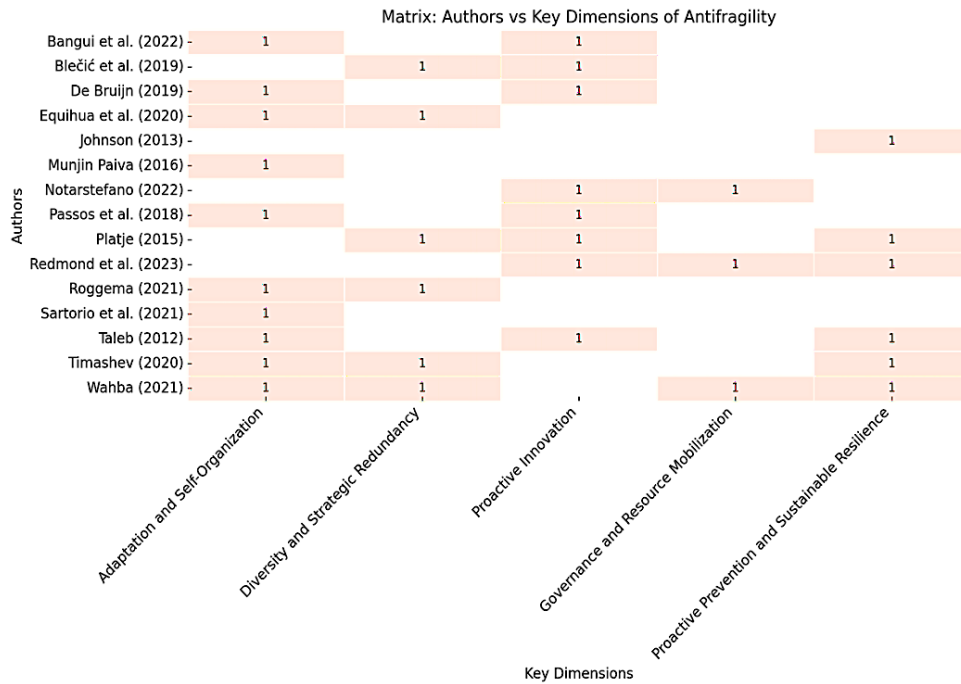

While this structuring approach offers a coherent reading grid, it remains too general to reveal the finer dynamics of the core concepts articulated by the 15 selected references. It risks obscuring weak signals and peripheral configurations that, though underexplored, hold significant transformative potential. To fully reflect the richness of these contributions, it is necessary to detail these dimensions by enriching them with principles illustrating the major concepts identified. These principles will form the basis of a theoretical framework that is both precise and operational. Table 2 elaborates the theoretical framework, where this mechanism of grouping the key concepts developed by the researchers (keys) are transcribed into unified (coded) and coherent fundamental theoretical principles.

From this exploration, we distill fifteen Fundamental Theoretical Principles of antifragility, that form the backbone of the urban antifragility framework proposed in this article. Conceived as a conceptual toolbox, the framework enables cities to treat disruptions as levers for innovation and continuous improvement. Rather than replacing resilience, it extends and reorients it by foregrounding the evolving dynamics of urban systems under crisis.

Principles whose aspects overlap or complement each other have been deliberately preserved, without ever duplicating each other. These slight overlapping effects are neither redundant nor incoherent, but a requirement for precision and complementarity. This structural choice aims to lay the foundations of an evolving fundamental conceptual framework, conceived as a starting point for an operationalized proposal.

*Findings: Antifragility Principles in Urban Post-Crisis Evolution*

To clarify the dynamics underpinning the framework summarized in Table 2, Figure 4 presents a conceptual map that analyses and synthesizes the distribution and relative importance of the 15 theoretical principles across the selected references. This mapping corroborates the framework's coherence and centrality by showing consistent activation across sources and isolating the core insights adopted in this research.

Figure 4 presents a matrix that cross-references the fifteen selected authors with the fifteen foundational principles of urban antifragility (A–P). Its analysis reveals dominant conceptual trends and exposes underlying dynamics essential to a deeper understanding of the framework.

*Table 2. Authors' panel and main contributions - Author: J.Uguet*

| Dimension | Key Concepts of the Authors | Emerging theoretical principle | Concise definition |
|---|---|---|---|
| **Adaptation & Self-organization** | *Continuous learning and self-organization* (dos Passos et al., 2018), *Adaptive urban morphology* (Sartorio et al., 2021) | **A. Learning from failure** | Turning mistakes into opportunities through iterative testing to improve urban responses. |
| | *Urban flexibility* (Blečić et al., 2020), *Morphology and adaptive modularity* (Sartorio et al., 2021) | **B. Systemic Flexibility** | Rapidly adapt urban policies and structures to disruptions, ensuring dynamic resilience. |
| | *Self-learning* (Notarstefano, 2022), *Collaborative decentralization* (dos Passos et al., 2018) | **C. Collaborative self-organization** | Develop decentralized systems that incorporate experience-based adjustments to strengthen their collective adaptability. |
| **Innovation Proactive** | *Blockchain and AI* (Bangui et al., 2022), *ICT and dynamical systems* (Passos et al., 2018) | **H. Reinforcement by chaos** | Harnessing crises to recombine and strengthen urban systems. |
| | *Innovative technologies* (Bangui et al., 2022), *Proactive anticipation of vulnerabilities* (Timashev, 2020) | **I. Proactive Innovation** | Introduce appropriate technologies and strategies to anticipate vulnerabilities and maximize operational impact. |
| | *Proactive Risk Prediction* (Timashev, 2020), *Predictive Models and Complex Scenarios* (Taleb, 2012) | **J. Systemic anticipation** | Predict emerging risks through forward-looking analyses and complex scenarios to prepare urban systems. |
| **Diversity & Strategic Redundancy** | *Structural redundancy* (De Bruijn, 2019), *Modularity* (Blečić et al., 2020) | **D. Positive redundancy and modularity** | Creating modular systems that can absorb shocks while ensuring continuity. |
| | *Strategic optionality* (Taleb, 2012), *Adaptability in the face of climate uncertainties* (Wahba, 2021) | **F. Strategic Diversity and Optionality** | Maintain a plurality of solutions to manage uncertainty and preserve strategic flexibility. |
| | *Distributed infrastructure* (Bangui et al., 2022), *Critical risk management* (Timashev, 2020) | **G. Decoupling of risks** | Partitioning critical systems to limit the propagation of failures. |
| **Governance & Resource Mobilization** | *Inclusive Governance* (Wahba, 2021), *Community Collaboration* (Notarstefano, 2022) | **K. Collective Participation** | Foster synergies between citizens, institutions and disciplines to integrate stakeholders into strategic decisions. |
| | *Rapid resource mobilization* (Redmond et al., 2023), *Effectiveness of deployments* (Timashev, 2020) | **L. Agile Engagement** | Rapidly deploy a variety of resources for immediate and effective responses to crises. |
| | *Interdisciplinary collaboration* (Wahba, 2021), *Social inclusion* (Notarstefano, 2022) | **M. Social Resilience** | Strengthen community networks for inclusive and sustainable collective responses. |
| **Proactive Prevention & Sustainable Resilience** | *Proactive sustainability* (Notarstefano, 2022), *Ecological approaches* (Equihua et al., 2020) | **N. Active prevention** | Identify and reduce vulnerabilities before they become problematic. |
| | *Value for money* (Notarstefano, 2022), *Sustainable ecological management* (Equihua et al., 2020) | **O. Sustainable resilience** | Develop systems aligned with circularity and optimization of natural resources. |
| | *Critical shock absorption* (Timashev, 2020), *Feedback* (Taleb, 2012) | **P. Shock absorption** | Maintain critical system functions during a crisis, incorporating feedback. |

*Figure 4. Intensity and Distribution by Author of the Theoretical Principles studied - Author: J.Uguet*

| Authors/Principles (codes) | A | B | C | D | F | G | H | I | J | K | L | M | N | O | P |
|---|---|---|---|---|---|---|---|---|---|---|---|---|---|---|---|
| Bangui et al. (2022) | | | | | | | 5 | 5 | 4 | | 2 | | | | |
| Blečić et al. (2019) | | 5 | | 4 | 4 | | | | 3 | | | | | 2 | 3 |
| de Bruijn (2019) | 4 | | | 5 | 3 | | | | 4 | | | | | | |
| Equihua et al. (2020) | | | | 4 | 3 | 3 | | | | | | | 4 | | |
| Johnson (2013) | 4 | | | | 3 | | | | 5 | | | 5 | | | |
| Munjin Paiva (2016) | 4 | 5 | 3 | | | | 3 | | | | | | | | |
| Notarstefano (2022) | | | 2 | | 3 | 4 | | | | 5 | | 4 | | 5 | |
| Passos et al. (2018) | 5 | | 5 | | 3 | | 3 | 4 | | | | 3 | | | |
| Platje (2015) | | | | 3 | 3 | | | | 3 | | | | 5 | 3 | |
| Redmond et al. (2023) | | | | | | | 4 | 5 | | 3 | | 3 | 3 | | |
| Roggema (2021) | | 3 | | 5 | 4 | 3 | | 3 | | | | | | | |
| Sartorio et al. (2021) | 3 | | | 5 | 3 | | | | | | | | | 3 | |
| Taleb (2012) | 5 | | | 5 | 5 | 4 | | 5 | | | | 4 | | | 4 |
| Timashev (2020) | | | 3 | | | | 5 | | 4 | | | 4 | | 5 | |
| Wahba (2021) | | | | | 4 | | | | | 5 | 3 | 3 | | 3 | 4 |
| | A | B | C | D | F | G | H | I | J | K | L | M | N | O | P |
| Total | 25 | 13 | 10 | 31 | 34 | 22 | 21 | 12 | 27 | 13 | 8 | 10 | 21 | 23 | 15 |
| % of Total 15 Principles | 9% | 5% | 3% | 11% | 12% | 8% | 7% | 4% | 9% | 5% | 3% | 3% | 7% | 8% | 5% |

Intensity / Role (0 to 5): 5 Main, 4 Important, 3 Moderate, 2 Low, 1 Indirect, 0 Null

| Letter | Principle |
|---|---|
| A | Learning from Failure |
| B | Systemic Flexibility |
| C | Collaborative Self-Organization |
| D | Positive Redundancy |
| F | Strategic Diversity |
| G | Risk Decoupling |
| H | Strengthening through Chaos |
| I | Proactive Innovation |
| J | Systemic Anticipation |
| K | Sustainable Governance |
| L | Agile Mobilization |
| M | Social Resilience |
| N | Active Prevention |
| O | Sustainable Resilience |
| P | Shock Absorption |

Rather than signalling consensus, these disparities point to methodological and epistemological diversity; nevertheless, they surface correlations, cross-cutting patterns, and telling absences that shape how scholarship articulates urban antifragility.

Certain principles emerge as indisputable "Pillars" of Urban Antifragility. Strategic Diversity, as shown in Column F, points to agreement that option-rich approaches most effectively reinforce adaptive capacity. This principle, central to Taleb (2012), De Bruijn (2019) and Roggema (2021), is closely linked to the Positive Redundancy in column D, which underlines the importance of modular and redundant structures. This F-D alliance forms a recurring conceptual basis for designing urban systems that can withstand disruption.

J *(Systemic Anticipation)* and O *(Sustainable Resilience),* together reflect an awareness of the need to prepare urban systems for prolonged shocks while adopting a long-term perspective. Johnson (2013) and Redmond (2023) emphasize the ability to predict and model crises, while Equihua (2020) and Sartorio (2021) place ecological sustainability at the centre of their thinking.

The analysis of the combinations reveals frequent alignments between certain principles. For example, A *(Learning through failure)* and F *(Strategic diversity)* are found together in several articles, reflecting a dynamic where proactive experimentation is supported by a diversity of adaptive solutions. This approach can be seen in dos Passos et al. (2018) and de Bruijn (2019), who put forward systems capable of transforming failures into opportunities.

Another notable correlation links J (Systemic Anticipation) and H (Reinforcement through Chaos), which are frequently co-activated, as in Bangui (2022) and Timashev (2020).

These studies advance a dual strategy: anticipate crises while harnessing disorder to spur innovation and reinforce existing structures.

Some principles, although essential in theory, appear to be under-represented. Principle I (Proactive Innovation), central to techno-structural frameworks such as those developed by Bangui (2022), is rarely mobilized. Similarly, Principle M (Social Resilience), emphasized by Wahba (2021) and Notarstefano (2022) for its role in community empowerment, remains marginal.

These shortcomings raise questions about a possible underestimation of these dimensions in current theoretical approaches. Variations in the activation of principles reflect distinct academic and methodological perspectives. Authors such as Taleb (2012) and De Bruijn (2019) take a systemic approach, valuing universal concepts such as redundancy and diversity. Conversely, contributions such as those of dos Passos (2018) and Bangui (2022) favour technological solutions, integrating blockchain, artificial intelligence and ICTs to strengthen urban innovation.

Others, such as Notarstefano (2022) and Wahba (2021), focus on governance and citizen participation, highlighting the importance of social dynamics in antifragility.

Finally, the matrix in Figure 4 corroborates the prominence of three foundational principles—F (Strategic Diversity), D (Positive Redundancy), and J (Systemic Anticipation)—that underpin urban antifragility. These are often reinforced through linkages to A (Learning from failure) and H (Reinforcement through Chaos). By contrast, I (Proactive Innovation) and M (Social Resilience), though underrepresented, may be pivotal in domain-specific or emerging contexts, especially technological or protracted crises.

Taken together, the mapping confers baseline legitimacy on all principles—even as some are theoretically dominant and others operate as context-dependent options conditioned by the adopted paradigm (e.g., Taleb vs. Dos Passos). It motivates future work on their interactions and synergies to bolster antifragile strategies in environmental, social, and technological contexts.

*Development of the Theoretical Principles of Urban Antifragility*

This section offers a concise exposition of the fifteen fundamental theoretical principles of urban antifragility, specifying their definitions and operational scope in post-crisis urban dynamics. Each principle is theoretically grounded and supported by contributions from the reference authors identified in our corpus.

Beyond the theory, these principles are part of a logic of operationalization "in fine", each one is illustrated by a concrete example of implementation, from documented experiences. These cases demonstrate how these dynamics, far from being abstractions, shape tangible and adaptable urban strategies. By confronting theory and application, this analysis aims to provide a pragmatic reading grid for decision-makers and practitioners, allowing them to identify the levers of urban transformation and to adapt these principles to specific contexts. This approach thus establishes a bridge between conceptualization and action, by integrating the theoretical and empirical dimensions of urban antifragility.

To support operationalization, Table 3 synthesizes the principles, specifying each code, a concise definition, and a concrete case example.

**Discussions and Conclusions**

*Pushing the Boundaries of Resilience: A Renewed Framework*

The analysis carried out in this paper highlights that while urban resilience is a stabilization strategy in the face of crises, it nevertheless has structural limits that hinder a sustainable evolution of urban systems. Traditional approaches tend to maintain the status quo rather than turn disruption into levers for improvement. Antifragility, based on the fifteen fundamental theoretical principles presented, goes beyond these limits by correcting institutional inertia through Systemic Flexibility (B) and Proactive Innovation (I), which allow for continuous adaptation, unlike resilience which generally freezes structures in rigid frameworks. Rather than absorbing shocks, it transforms them into opportunities for evolution through Chaos Reinforcement (H) and Proactive Experimentation (A).

Where resilience comes up against a compartmentalized approach, antifragility offers Strategic Diversity (F) and Systemic Anticipation (J) that ensure a plurality of solutions in the face of uncertainty. By building local and collaborative capacities through Collective Participation (K) and Agile Mobilization (L), it actively mobilizes its resources and integrates multiple actors in an inclusive and adaptive model.

Finally, by promoting functional autonomy with Risk Decoupling (G) and Modularity (D), it guarantees decentralized management, thus avoiding the domino effects of crises. These elements frame antifragility as an operational model that complements resilience—sidestepping common pitfalls and enhancing cities' long-term adaptive capacity.

| Code | Theoretical principle | Explanatory synthesis of the concept | Suggested example |
|---|---|---|---|
| A | Learning from failure & proactive experimentation | Exploitation of errors through iterative testing in controlled environments to optimize urban systems and anticipate crises. Promotes incremental innovation and adaptive sustainability through continuous, low-risk adjustments (Taleb, 2012; Passos, 2018; Johnson, 2013). | Singapore, Smart Nation initiative (proactive water management via IoT). |
| B | Systemic & institutional flexibility | Ability to adapt urban structures and processes to disturbances, through modular infrastructures, dynamic governance and experiments in a controlled context (Blečić et al., 2020; Paiva, 2016). | Rotterdam, "Living with Water" strategy. |
| C | Collaborative self-organization & self-learning | Reduction of central dependence through decentralized coordination inspired by stigmergy (Grassé, 1959), promoting autonomy and collective adaptation in the face of complex crises (Passos et al., 2018; Johnson, 2013). | Barcelona, COVID-19 Community Networks (Xarxa d'Ajuda Mútua). |
| D | Positive redundancy & modularity | Diversification of urban systems to avoid single breaking points through the adoption of modular and adaptive solutions, ensuring operational continuity (Roggema, 2021; Blečić et al., 2020). | Rotterdam, "Maeslantkering" flood barriers. |
| F | Strategic diversity & optionality | Proactive integration of multiple and complementary solutions, allowing flexibility and adaptability in the face of complex crises and uncertainties (Taleb, 2012; Bangui et al., 2022). | Tokyo, earthquake-resistant infrastructure and microgrids. |
| G | Decoupling risks | Limitation of interdependencies between systems to prevent domino effects in the event of failure, promoting autonomy and robustness of critical infrastructures through distributed approaches (Taleb, 2012; Bangui, 2022). | Tel Aviv, energy microgrids and autonomous systems. |
| H | Reinforcement by chaos | Use of disturbances as levers for innovation and improvement, generating cycles of structural and functional recombination of complex systems (Taleb, 2012; Redmond et al., 2023). | Fukushima, post-disaster energy innovations. |
| I | Proactive innovation | Risk anticipation through the integration of advanced technologies (IoT, AI) to reduce vulnerabilities and optimize urban interventions (Bangui et al., 2022; Passos et al., 2018). | Singapore, Smart Nation (IoT sensors, AI). |
| J | Systemic anticipation | Forecasting emerging risks via complex scenarios and advanced modelling (AI-based) to optimise urban crisis management (Bangui, 2022; Johnson, 2013). | Paris, "Ile-de-France flood strategy (2023)". |
| K | Collective participation & interdisciplinary collaboration | Active stakeholder engagement through inclusive governance and interdisciplinary synergies to strengthen crisis responses (Wahba, 2021; Notarstefano, 2022). | Medellín, Proyectos Urbanos Integrales (PUI). |
| L | Agile Engagement | Rapid and flexible activation of resources to respond to crises, supported by decentralized coordination and innovative technologies (Blockchain, AI) (Timashev, 2020; Redmond et al., 2023). | New Orleans, post-Katrina management. |
| M | Social resilience | Strengthening community autonomy and local networks for an inclusive and responsive response to crises, optimizing social cohesion (Notarstefano, 2022; Wahba, 2021). | Barcelona, Social Resilience Plan 2021-2030. |
| N | Active prevention | Proactively reducing vulnerabilities through regular audits and predictive methodologies, supporting continuous adaptation to emerging threats (Redmond et al., 2023; Platje, 2015). | Tel Aviv, Home Front Command (proactive readiness). |
| O | Sustainable resilience | Sustainable optimization of natural resources through circularity and scalability of infrastructure, integrating nature-based solutions (NBS) (Equihua et al., 2020; Notarstefano, 2022). | Jinan (China), Sponge City concept. |
| P | Shock absorption | Maintaining critical functions of systems in times of crisis with continuous integration of lessons learned to strengthen their robustness (Taleb, 2012; Timashev, 2020). | Tokyo, anti-seismic infrastructure and anti-seismic systems |

*Table 3 - Framework of Fundamental Theoretical Principles of Urban Antifragility*

*Framing Post-Crisis Urban Antifragility: A Theoretical Basis for Operationalization*

Urban antifragility reframes crisis not as a disturbance to absorb, but as a catalyst for structural evolution. Far beyond resilience, it installs cities within a logic of continuous learning, where each shock becomes a trigger for adaptation and transformation. The challenge now lies in translating this vision into an operational framework capable of assessing, enhancing, and guiding the capacity of cities to thrive under constraint.

While urban antifragility is a major conceptual evolution, its implementation poses several challenges. Rigid institutional frameworks remain a major obstacle, with traditional governance mechanisms favouring stability rather than transformation, making it difficult to adopt an approach based on experimentation and adaptability. In addition, its integration into public policies requires an overhaul of urban planning tools and the creation of regulatory frameworks that facilitate innovation and flexibility. Moreover, the absence of clear indicators for urban antifragility underscores the need to develop metrics that can rigorously measure the impact of implemented strategies. Despite this gap, antifragility offers notable strategic advantages—chief among them the convergence of smart technologies and adaptive systems that enable rapid, scalable crisis responses. In parallel, urban models already embedding antifragile principles are emerging, providing concrete testbeds for experimentation and evaluation

Therefore to operationalize this framework four strategic axes emerge as summarized in Figure 5. First, a systematic empirical analysis of urban contexts where antifragile logics have been activated—whether intentionally or not—will provide concrete insights into the mechanisms of post-crisis transformation. By examining institutional responses, socio-spatial shifts, and structural innovations, this axis seeks to test the relevance and activation of the fifteen theoretical principles in real-world conditions.

Second, the conceptual framework must be operationalized into a structured toolkit (to assess, plan, and monitor). This involves translating the principles into strategic dimensions, attributes, and measurable indicators, informed by empirical feedback. The objective is to offer decision-makers a usable structure—flexible, scalable, and adapted to multi-level governance, from strategic design to implementation. In this toolkit, a curated compendium of best practices will operate as a dynamic repository of validated interventions. Organized by activated principles and associated performance metrics, it will guide context-sensitive recommendations—for example, Bogotá's TransMilenio (redundancy and social inclusion), Barcelona's Superblocks (spatial flexibility), and Rotterdam's Water Squares (adaptive climate response).

Third and finally, empirical validation through application in cities exposed to multi-layered crises will allow for iterative testing and refinement of both the framework and its toolkit.

Together, these three axes constitute the backbone of a validation program grounded in comparative urban case studies. By systematically confronting the model with urban realities, the aim is to iteratively refine the framework, embed it in practice, and make antifragility a usable lever for sustainable, adaptive, and scalable urban transformation.

Operationally, this entails mixed-method triangulation (document analysis, spatial–temporal indicators, and stakeholder interviews), cross-case synthesis, and targeted pilot implementations with monitoring dashboards—yielding principle-to-action playbooks, context profiles, and metrics that track performance, equity, and learning.

Figure 5 synthesizes the operationalization process, including the first step which has been developed in this paper.

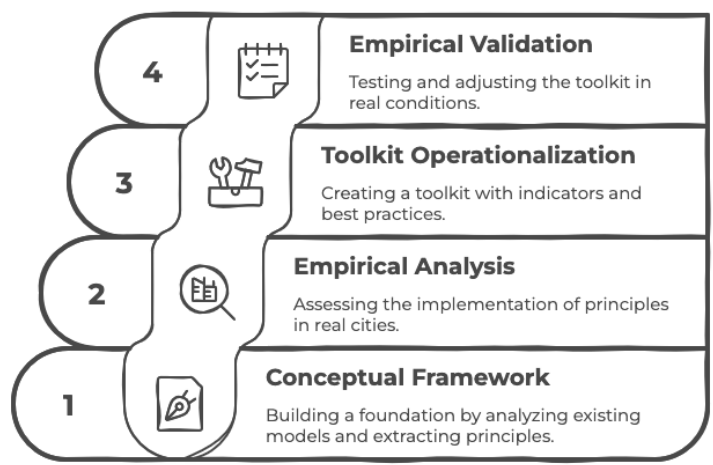

*Figure 5. Developing an Operational Urban Antifragility Framework. Author: J.Uguet*

**Conclusion**

Urban crises reveal more than vulnerabilities, they expose dormant potentials for transformation. This article has laid the groundwork for a theoretical shift by articulating fifteen foundational principles of antifragility, drawn from a critical synthesis of cross-disciplinary research. By mapping these principles and confronting them with empirical trajectories, it distinguishes antifragility from resilience, not as a refinement, but as a paradigm change: from recovery to regeneration, from adaptation to evolution. The proposed "toolbox" consolidates these insights into a coherent theoretical foundation, designed to guide urban systems in navigating uncertainty, learning from shocks, and activating endogenous dynamics of change. Its architecture is deliberately open, allowing for contextual interpretation and future enrichment. As cities face intensifying disruptions, this approach offers more than a conceptual lens, it becomes a lever for strategic reorientation. Anchored in theory yet oriented toward practice, it invites a reframing of sustainable urban futures: not as systems that endure, but that evolve, precisely because they are exposed to disorder.

**Appendix 1: Summary list of the main contributions to antifragility: 33 authors (non-exhaustive list)**

| No | Author(s) & date | Contribution Area | Main themes covered | Key Contributions |
|---|---|---|---|---|
| 1 | **Axenie (2023)** | Urban systems | Complexity & Adaptive Systems, Innovation & Urban Foresight | Analysis of the antifragility of complex systems in the face of disturbances and time. |
| 2 | **Bangui (2022)** | Urban governance | Adaptive governance / institutions, Anticipation & risk management | Study of the application of antifragility to critical infrastructure in the digital age. |
| 3 | **Bendell (2014)** | Theoretical foundations | Urban Sociology & Social Capital, Adaptive Governance / Institutions | Exploring the ethical aspects of intelligent and adaptive systems. |
| 4 | **Blečić (2020)** | Urban ecosystems | Innovation & Urban Foresight, Ecosystem Approaches | Application of antifragility concepts in urban planning. |
| 5 | **Corvello (2022)** | Urban governance | Participation & collective intelligence, Innovation & urban foresight | Integration of antifragility principles into organizational innovation models. |
| 6 | **Danchin (2011)** | Urban ecosystems | Ecosystem Approaches, Innovation & Urban Foresight | Application of antifragility concepts to biological and organizational systems. |
| 7 | **Davoudi (2012)** | Theoretical foundations | Adaptive Governance / Institutions, Complexity & Adaptive Systems | Analysis of urban resilience and foundations for a reflection on antifragility. |
| 8 | **de Bruijn (2019)** | Urban systems | Redundancy, modularity & diversification, Anticipation & risk management | Deepening redundancy and progressive experimentation strategies to limit systemic fragilities. |
| 9 | **De Florio (2014)** | Urban systems | Complexity & Adaptive Systems, Redundancy, Modularity & Diversification | Proposal of a model combining elasticity, resilience and machine learning. |
| 10 | **Equihua (2020)** | Urban ecosystems | Ecosystem Approaches, Ecological Transition / Sustainability | Study of the application of antifragility to urban ecosystems and biodiversity. |
| 11 | **Grassé (1959)** | Theoretical foundations | Ecosystem Approaches, Participation & Collective Intelligence | Introduction of the notion of stigmergy, the basis of self-organization in complex systems. |
| 12 | **Hill (2020)** | Theoretical foundations | Anticipation & risk management, Innovation & urban foresight | Interdisciplinary research on the application of antifragility. |
| 13 | **Holling (1973)** | Theoretical foundations | Ecosystem Approaches, Redundancy, Modularity & Diversification | Founder of the concept of ecological resilience, essential to understanding the evolution of systems. |
| 14 | **Jacobs (1961)** | Urban governance | Urban Sociology & Social Capital, Participation & Collective Intelligence | Pioneering vision of urban diversity and the organic development of cities. |
| 15 | **Johnson (2013)** | Urban systems | Anticipation & risk management, Complexity & adaptive systems | Development of a methodological framework to analyse and measure antifragility in complex systems. |
| 16 | **Karadimas (2014)** | Urban governance | Adaptive Governance / Institutions, Redundancy, Modularity & Diversification | Application of Taleb's theories to resilience cycles in virtual organizations. |
| 17 | **Lamarck (1809)** | Theoretical foundations | Complexity & Adaptive Systems, Ecosystem Approaches | Proposal of the idea of evolutionary transformation through environmental adversity, prefiguring concepts related to antifragility. |
| 18 | **Liu (2019)** | Urban systems | Complexity & Adaptive Systems, Innovation & Urban Foresight | Analysis of antifragility in critical and adaptive infrastructure. |
| 19 | **Meerow et al. (2016)** | Urban ecosystems | Redundancy, Modularity & Diversification, Ecosystem Approaches | Approach integrating diversity and modularity in complex urban systems to promote antifragility. |
| 20 | **Munjin Paiva (2016)** | Urban ecosystems | Ecosystem Approaches, Ecological Transition / Sustainability | Exploration of the application of antifragility to architecture and built heritage. |
| 21 | **Munoz (2022)** | Theoretical foundations | Redundancy, Modularity & Diversification, Innovation & Urban Foresight | Conceptual differentiation between robustness, resilience and antifragility. |
| 22 | **Notarstefano (2022)** | Urban governance | Ecological transition / sustainability, Adaptive governance / institutions | Study of the relationship between sustainability and antifragility in the context of the Sustainable Development Goals. |
| 23 | **Dos Passos et al. (2018)** | Urban systems | Innovation & Urban Foresight, Ecosystem Approaches | Transition from resilience to antifragility through information technology. |
| 24 | **Platje (2015)** | Theoretical foundations | Adaptive governance / institutions, Anticipation & risk management | Consideration of fragilities and antifragility in the context of sustainable development. |
| 25 | **Redmond et al. (2023)** | Urban governance | Anticipation & risk management, Innovation & urban foresight | Complex modeling and antifragile management of organizational systems. |
| 26 | **Roggema (2021)** | Urban systems | Redundancy, Modularity & Diversification, Innovation & Urban Foresight | Analysis of the role of spatial redundancy in improving the adaptive capacity of cities. |
| 27 | **Sartorio et al. (2021)** | Urban systems | Innovation & Urban Foresight, Adaptive Governance / Institutions | Exploration of urban planning models integrating antifragility through experimental approaches. |
| 28 | **Shafique (2016)** | Urban systems | Innovation & Urban Foresight, Adaptive Governance / Institutions | Study of urban antifragility as a transformative alternative to resilience. |
| 29 | **Taleb (2012)** | Theoretical foundations | Anticipation & risk management, Complexity & adaptive systems | Introduction of the concept of antifragility, defining systems that strengthen in the face of shocks and uncertainty. |
| 30 | **Terje Aven (2015)** | Theoretical foundations | Anticipation & risk management, Adaptive governance / institutions | Analysis of risk and antifragility as complementary paradigms. |
| 31 | **Timashev (2020)** | Urban systems | Anticipation & risk management, Redundancy, modularity & diversification | Concept of supra-resilience of urban critical infrastructure in the face of unforeseen disasters. |
| 32 | **Tokalić et al. (2021)** | Urban governance | Adaptive governance / institutions, Ecological transition / sustainability | Integration of antifragility into innovation and governance strategies. |
| 33 | **Wahba (2021)** | Case Studies | Adaptive governance / institutions, Participation & collective intelligence | Exploring people-and-place approaches for decentralized self-organization in post-conflict contexts. |

24